# Tailoring the Hydrophobicity of Mesoporous Organosilica for Protein Trapping and Supported Catalysis.

Oriana Osta*[a], Marianne Bombled[a], David Partouche[a,c], Florian Gallier[b], Nadège Lubin-Germain[b], Nancy Brodie-Linder*[a,b], and Christiane Alba-Simionesco*[a].

[a] *Laboratoire Léon Brillouin, CEA Saclay, Bâtiment 563, 91191 Gif-sur-Yvette, France.*

[b] *Laboratoire de Chimie Biologique, Université de Cergy-Pontoise, 5 mail Gay-Lussac, Neuville sur Oise, 95301 Cergy-Pontoise Cedex, France.*

[c] *Synchrotron SOLEIL, L'Orme des Merisiers, Saint Aubin, 91192 Gif-sur-Yvette, France.*

*Supporting Information Placeholder*

**ABSTRACT:** We propose a method to enhance lysozyme trapping and supported-Copper catalysis when confined in organosilica materials. The direct synthesis presented here allows the control of the silica surface hydrophobicity by uniform introduction of methyl or phenyl groups. As a result, the lysozyme trapping is observed to be 3.2 times more efficient with the phenyl-functionalized material than MCM-41. For heterogeneous catalysis, Copper was immobilized on the new organosilica surface. In this case, the presence of methyl groups significantly enhanced the product yield for the catalyzed synthesis of a triazole derivative. This method opens a new route of synthesis where the material properties can be adjusted and dedicated to a specific application.

## I. Introduction

Molecular trapping and supported catalysis are applications of mesoporous materials with great industrial and scientific impact. A key parameter for their efficiency is the hydrophobicity of the surface, accompanied by a large surface area and the appropriate pore size[1]. Classic examples for applications in molecular trapping are the encapsulation followed by controlled drug delivery in nanomedicine[2], the effective removal of the low concentrations of pollutants[3] and molecular confinement studies[1,4–6] to mention a few. In the area of catalysis, mesoporous silica-supported reactions have been investigated for example, in wastewater treatment[7], in reducing the amount of certain compounds in the mainstream of tobacco smoke[8] and in product yield improvement for organic reactions,[9–11] among other applications. In a particular case, copper doped nanoporous silica have been found to be a reusable, regioselective catalyst for the cycloaddition reaction between a terminal alkyne and an azide through click chemistry.[10,12] However, in some cases, the yields of reaction are very low and this is attributed to the low interaction between the hydrophobic reactants and the hydrophilic materials.

In this context, MCM-41 materials offer high surface area, ordered structure and thermal stability, but it fails in providing the correct hydrophobic/hydrophilic balance for an efficient interaction with most organic molecules. Frequently, even after post-grafting hydrophobic groups on the surface, this balance is not met due to a total transformation of the surface from hydrophilic to hydrophobic. A partial functionalization with this method has shown to produce inhomogeneous, rough surfaces with the majority of the grafted groups on the external surface or at the extremes or the pores.[9,13,14]

The co-condensation method has been conceived as an alternative to introduce hydrophobic groups to mesoporous silica materials yielding to a more uniform functionalization[9,15]. In this case, a silica precursor is condensed together with an organosilane. This method has been studied to introduce acid, basic and unsaturated groups in the walls of MCM-41-like materials so that they can be directly used for catalytic applications.[14,16–18] Our aim was to control the hydrophobicity of organosilica materials for specific applications by introducing methyl or phenyl groups through co-condensation with methyltriethoxysilane (MTES) or phenyltriethoxysilane (PTES). Such an approach for the synthesis of organosilica molecular sieves[19] has been much less investigated.

Here, lysozyme was chosen as a model to study the effect of the surface hydrophobicity in protein uptake. In addition, the supported copper (click) catalysis of the cycloaddition reaction between two hydrophobic reactants, benzyl azide and 4-bromo-1-butyne, in aqueous media instead of an organic solvent, was tested with the synthesized materials.

For the direct synthesis of mesoporous organosilica, we focused on obtaining a stable material with a highly ordered pore structure. In order to avoid the disruption of the honeycomb-like structure and based on previous research,[20] the percentage of organosilane in the precursor mixture was kept under 20 mol%. The conditions of template washing by ethanol extraction were optimized to obtain a clean material without causing damages to the surface.

## II. Experimental Section

**Materials**

Octadecyltrimethylammonium bromide (C21H6NBr), Dodecyltrimethylammonium bromide (C15H6NBr), Tetraethyl orthosilicate (TEOS), methyltriethoxysilane (MTES), phenyl-triehoxysi-lane (PTES), Ammonia (30%), Hydrochloric acid (37%) and Ethanol (99.98%) were purchased from Sigma-Aldrich.

**Direct Synthesis of modified MCM-41.**

MCM-41-type materials with different percentages of MTES or PTES were prepared. For each one, 4 mmol of the cationic surfactant CnTAB was dissolved in 72.0 ml of ultrapure water at 50 °C and was stirred vigorously during 1.5 h for complete dissolution, then the temperature was decreased to 35 °C and 4.8 ml of ammonia (30%) were added, the mixture was stirred for fur-ther 5 min. After, the precursors' mixture was added keeping a constant number of moles of Silicon (26.33 mmol). The final molar composition of the reaction mixture was (1 – x) TEOS: x MTES (or PTES): 2.4 NH3: 0.15 CnTAB: 152 H2O where x is the fraction of methyl- or phenyl-triethoxysilane that goes from 0 to 0.15 and n is the length of the alkyl chain of the surfactant. The reaction was allowed stirring, at rate of 300 rpm, during 3.5 h. The obtained white solid was washed twice with 40 ml of HCl 2%v/v and then with ultrapure water until reaching neutral pH.

Samples were named according to the template used and their fraction of MTES or PTES in the precursor mixture, as follows: C18-M7-Me for 0.93 TEOS/ 0.07 MTES, C18-M15-Me for 0.85 TEOS/ 0.15 MTES, C18-M7-Ph for 0.93 TEOS/ 0.07 PTES, all of them using $C_{18}$TAB as template, and C12-M7-Me for 0.93 TEOS/ 0.07 MTES using $C_{12}$TAB as template. Standard MCM-41 C18 and C12 were synthesized and washed under the same conditions for comparison.

For template removal, 100 ml of a solution of HCl (2%v/v) were added per gram of product and it was stirred for 25 min at room temperature. The solid was filtered and re-dissolved in a mixture of 200 ml of Ethanol 99% and 10ml of ultrapure water. This mixture was stirred at 50 °C during 3 h, then, the solid was filtered and dried in the oven at 60 °C during 24 h. Only for sample C18-M7-Ph an extra thermal treatment at 350 °C during one hour under nitrogen atmos-phere was needed for complete template removal.

**Characterization**

FTIR spectra of the samples were taken from 4000 to 400 cm-1, with a resolution of 2 cm-1, with a Bruker TENSOR equipped with a Platinum ATR. Nitrogen and water vapor adsorption-desorption experiments were performed with ASAP 2020 Surface Area and Porosity Analyzer from micromeritics. Nanoporous silica samples were degassed under vacuum at 90 °C during 24 h. Nitrogen adsorption isotherms were taken at 77 K, and water vapor adsorption isotherms at 298.15 K. Thermogravimetric analyses were performed with a Q50 thermoanalyzer from TA In-struments, with controlled atmosphere under nitrogen flow, from ambient temperature to 850 °C with a temperature ramp of 10 °C/min. Small-angle X-ray Scattering patterns were collected using a Xeuss 2.0 HR SAXS/WAXS instrument from Xenocs with a Cu Kα (5 kV, 0.6 A). TEM Images were taken with a JEOL JEM-2010 high resolution transmission electron microscope, operating at 200 kV.

**Lysozyme trapping**

Approx. 10 mg of the selected materials were mixed with 10 ml of a 0.5 mg/ml solution of lysozyme from chicken egg white, in a 10 mM phosphate buffer pH 7.4, during 4 h at room temper-ature. Then, the suspensions were centrifuged during 10 min at 15000 x g, 1 ml of the superna-tant was passed through a 0.2 μm filter and transferred to a quartz cell. The UV absorption was taken at 280 nm. The UV absorption spectra of the rinsing water were also taken to discard the loss of weakly bounded lysozyme. The solids were rinsed with water and dried under vacuum, the FTIR spectra of the lysozyme-rich materials was taken.

**Copper post-grafting**

The addition of copper followed the protocol described by Nancy Brodie-Linder et al1. 150 mg of the washed nanoporous silica materials were added to 40 ml of a 0.05 M solution of copper nitrate, previously adjusted to pH 10.5 with ammonia (30%), and stirred vigorously for 10 min. Immediately after, the solid was filtered and washed with ultrapure water until neutral pH. The obtained blue material was dried in the oven at 50 °C. The abovementioned procedure was also applied on the as synthesized samples (without removing the surfactant from the pores), in or-der to obtain materials with copper grafted only on the external surface.

**Catalytic tests**

The protocol for the catalysis experiments of the click reaction using copper doped nanopo-rous silica was the same as published before2. For all samples, Benzyl azide (67 μl) and 4-bromo-1-butyne (47 μl) were added to 2 ml of water con-taining the 8 mol% of copper supported in the different silica materials. Reactions were stirred during 22 h. After that time, the mixtures were vacuum filtered with a polyacryla-mide membrane and the blue solids were rinsed twice with 2ml of ethyl acetate each time. After solvent evaporation, the conversion to 1-benzyl-4-bromoethyl-triazole was deter-mined by 1H NMR.

**Neutron Backscattering Experiments**

The backscattering spectrometer IN16B allows experiments with a the high-energy-resolution ΔE = 0.75 μeV at FWHM (equivalent to 5 ns) with a neutrons wavelength λ = 6.3 Å, in a wave vector range of 0.2 to 1.9 Å -1, however the data pre-sented are the sum of the Q scans obtained at each tempera-ture. Standard Si(111) monochromator and analyzer crystal setup were used here. The experiments were operated in two different modes: we have measured the scattered intensi-ty in the elastic fixed window (EFWS) mode and in the inelastic fixed window mode (IFWS) at energy of 2 eV and performed several temperature scans from 2K to 290K. A flat Aluminum cell was used and filled with the 2 materials, MCM-C18-M15-Me and MCM-C18-M15-Me partly calci-nated, in a glove box to avoid water contamination. The samples were first rapidly cooled and then heated at a rate of 2K/min until 100K and at 0.8K/min from 100K to 290K to improve the statistics.

**III. Results**

Template removal and incorporation of the organic groups in the mesoporous silica was verified by means of FTIR (Figure S1 in ESI). Nitrogen adsorption and desorption isotherms at 77 K for these samples are shown in Figure 1. For standard and methyl modified MCM-41 C18, results show a regular type IV adsorption isotherm. In the case of co-condensation of TEOS with MTES, it was observed an increase of surface area and pore volume with respect to MCM-41 C18 (See Table 1). On the other hand, the co-condensation with PTES pro-duced a less organized material, evidenced by a spread capil-lary condensation, slightly lower surface area and pore volume than MCM-41 C18.

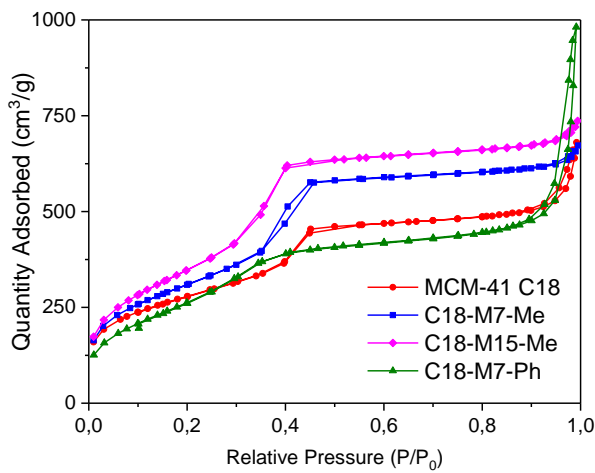

**Figure 1.** Nitrogen adsorption-desorption isotherms at 77 K of MCM-41 C18 and mesoporous organosilica materials.

The late condensation at $P/P_0$ higher than 0.9 in samples MCM-41 C18 and C18-M7-Ph was attributed to condensation in the interparticle cavities when the size of the mesoporous silica grains are in the nanometer scale[21], as it was evidenced for C18-M7-Ph by TEM (Figure S3).

These results indicate that upon introduction of MTES at percentages up to 15%, the obtained organosilica maintain an ordered pore structure while increasing their hydrophobic character. To investigate the upper limit of the fraction of MTES in the precursors mixture that could yield an ordered, highly porous material, an extra sample was synthesized with 20% MTES. Results show that even though a high BET surface area (903 m$^2$/g) is obtained, there is a significant decrease in the pore volume (0.486 cm$^3$/g) of the material.

Likewise, sample C12-M7-Me was synthesized in order to explore the hydrophobic modification of small pore mesoporous silica, which is difficult to obtain by post-grafting methods. In this case, a high surface area material, with pore volume slightly lower than the MCM-41 C12 was obtained. The production of such small pore sized material with controlled hydrophobicity, like the one obtained in this work, could be of great scientific interest for molecular confinement studies.

It is worth noting that the affinity constant, C, calculated from the BET equation, decreased 31%, 49% and 61% with respect to the value of MCM-41 C18 (C = 112) for C18-M7-Me, C18-M15-Me and C18-M7-Ph, respectively. This parameter decreases with a decrease on the strength of the adsorbent-adsorbate affinity. Since induced dipole-dipole interactions govern this adsorption, the introduction of organic groups decreases the affinity of nitrogen towards the surface. Then, the observed drop of the C value within the same family of materials is another evidence of their increasing hydrophobicity.

Figure 2 shows the Small-angle X-Ray scattering patterns for the synthesized materials. The observation of three Bragg peaks indexed as (10), (11) and (20) confirm the 2D hexagonal structure p6mm of the standard, and methyl modified materials. The broadening of the (10) peak of the phenyl modified sample could suggest a higher polydispersity in the pore size.

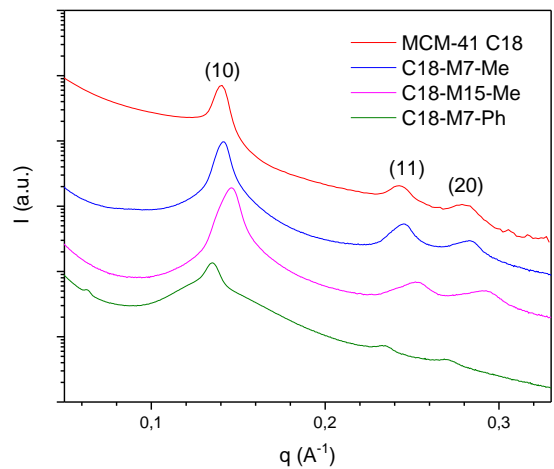

**Figure 2.** Small angle scattering curves of MCM-41 C18 and mesoporous organosilica materials.

The lattice plane spacing $d_{100}$ was calculated from the SAXS patterns and it is shown in Table 1. This parameter in combination with the pore volume obtained by nitrogen adsorption allowed the calculation of the pore diameter ($D_{pore}$). This method has been found to be more precise for the determination of the size of small hydrophobic pores than other models that use exclusively nitrogen adsorption data[22]. We found that up to 15% MTES or 7% PTES, the pore diameter does not change significantly with respect to MCM-41, while the post graphing method induces always a decreasing and less predictable size.[22]

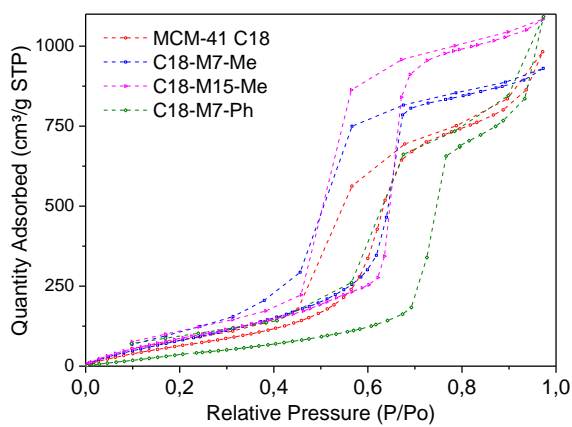

**Figure 3.** Water adsorption isotherms at 298.15 K of (a) MCM-41 C18, (b) C18-M7-Me, (c) C18-M15-Me and (d) C18-M7-Ph.

**Table 1.** Mean physical properties of the mesoporous organosilica materials.

| Sample | $S_{BET}$ (m²/g) | $V_{pore}$ (cm³/g)[a] | $d_{10}$ (nm) | $D_{pore}$ (nm)[d] | Cu(II) (%w/w) |
|---|---|---|---|---|---|
| MCM-41 C18 | 1007 | 0.712 | 4.45 | 4.22 | - |
| C18-M7-Me | 1130 | 0.899 | 4.48 | 4.43 | - |
| C18-M15-Me | 1284 | 0.982 | 4.31 | 4.32 | - |
| C18-M20-Me | 903 | 0.447 | 4.05 | 3.46 | - |
| C18-M7-Ph | 989 | 0.628 | 4.65 | 4.30 | - |
| MCM-41 C12 | 979 | 0.539 | 3.55 | 3.17 | - |
| C12-M7-Me | 1134 | 0.505 | 3.37 | 2.97 | - |
| Cu/MCM-41 C18 | 308 | 0.286[b] | [c] | - | 14.5 |
| Cu/C18-M7-Me | 256 | 0.181 | 4.46 | 2.89 | 9.8 |
| Cu/C18-M15-Me | 377 | 0.227 | 4.32 | 3.02 | 9.9 |
| Cu/C18-M7-Ph | 430 | 0.331 | 4.55 | 3.58 | 9.9 |

[a] Single point adsorption total pore volume at $P/P_o = 0.5$ [b] Single point adsorption total pore volume at $P/P_o = 0.8$ [c] Not measured [d] Calculated using the KJS model taking into account the lattice plane spacing ($d_{10}$).[23]

Water adsorption isotherms (Figure 3) show that the total quantity of water increases when the pore volume of the material is higher, regardless of their organic functionalization These results show that the hydrophobicity of the materials is not strong enough to avoid water vapor condensation in the pores.

The stability of the materials towards hydrolysis was studied by performing nitrogen adsorption analysis before and after of water adsorption. Results (Figure S2) clearly show that samples C18-M15-Me and C18-M7-Ph are better stabilized, with a smaller loss of pore volume than C18-M7-Me. The water-resistance found here, could be crucial for better performance of the molecular sieves in aqueous solution.

Transmission electron microscopy images of selected samples are shown in Figure S3. Both C18-M7-Me and C18-M7-Ph show the highly ordered, hexagonal structure, surrounded by mesocellular silica foam (MCF), reported before when 1,3,5-trimethylbenzene was incorporated to the synthesis of the mesoporous silica[24]. A possible explanation is that MCF rises due to the formation of a microemulsion upon addition of the organosilane, which behaves as a swelling agent, due to its hydrophobic character. The abundance of MCF structures in C18-M7-Ph explains the increased width of the Bragg's peaks in the SAXS pattern for this sample.

Additionally, in order to verify the presence of methyl groups at the surface of the pore, we have conducted quasielastic neutron scattering experiments in the fixed elastic mode window mode of 2 μeV. From the T-dependent of the elastic scans obtained, we could identify a pronounced enhanced mobility of the methyl groups starting at 80K.

We have investigated the presence of methyl groups at the surface and in the wall of MCM-C18-M15-Me and distinguished the onset of their free rotation about the threefold axis as temperature increases from 2K to 290K. For that purpose Incoherent Neutron Backscattering method is particularly well suited because of its high sensitivity to Hydrogen atoms and its possibility to operate Elastic and Inelastic Fixed Window temperature Scans, EFWS and IFWS, which can easily identify the onset of motions faster than few nanoseconds. With the high resolution backscattering spectrometer IN16B at the ILL (Grenoble, France)[25], we have performed two sets of experiments[26,27] on two different samples. The first sample is the C18-M15-Me, compared to its partly calcined homologue, the latter being heated up to 550°C in order to suppress a large part of the methyl groups attached at the surface. The results are presented in Figure 4 (a, b) where respectively the two EFWS normalized at 2K and the two IFWS and their intensity difference are reported.

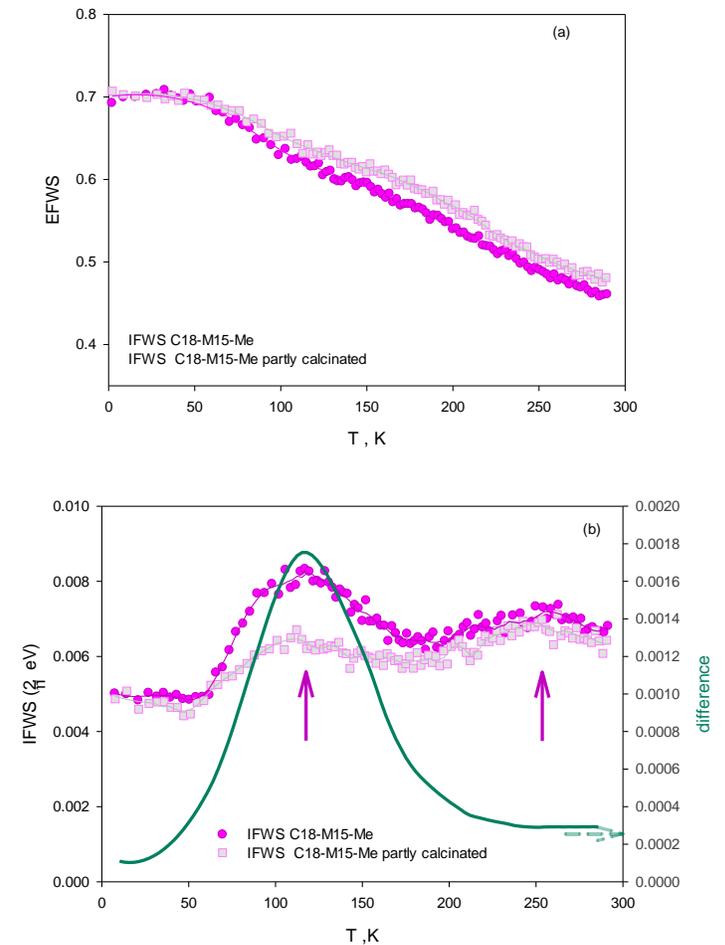

**Figure 4:** (a) Elastic intensity measured with the fixed window mode, EFWS, as a function of temperature on the spectrometer IN16B at the ILL, for the two samples MCM-C18-M15-Me, as defined in the main text, and its partly calcined

homologue (C18-M15-Me partly calcined); (b) Inelastic intensity measured in the IFWS mode at 2µeV; pink arrows indicate the temperatures at which the methyl group rotations cross the energy of 2µeV, the difference between the 2 samples in green illustrates how the methyl groups at the surface are affected by the thermal treatment.

At very low temperature all motions of the materials are frozen-in and the elastic scattered intensity slightly decreases with temperature because of Debye-Waller factor; however local motions such as methyl group rotation can take place[27] and the EFWS in Figure 4a decays in two steps indicating that above 100 K and 250 K the methyl group rotation enters the dynamical window of the spectrometer. The 2 steps-EFWS demonstrates that the methyl groups are interacting with two distinct environments: the less constraining one is at low temperature and corresponds to the groups at the surface, while the most constraining one is at higher temperatures and corresponds to groups in the wall; the slight difference in intensity above 100K is due to the most rigid character of the calcined sample. These results are confirmed in Figure 4b with the IFWS; as the temperature increases, the frequency of the rotation shifts to lower value and enters in the elastic line, causing a quasi-elastic broadening. Detected at a fixed frequency of 2 µeV, the inelastic signal increases in amplitude; as T further increases, the motion becomes faster and the inelastic signal even broader, inducing an amplitude decrease at fixed frequency. Two maxima are seen in the IFWS corresponding to the 2 steps in EFWS and indicated by the arrows in the figure. At 117 K, the difference in amplitude between the 2 samples evidences the presence of the groups at the surface that are affected by the thermal treatment of the calcined one. But the second important feature is the lowest inelastic intensity of the peak at 255 K illustrating the lowest amount of mobile H in the walls, which remain the same in the two samples.

**Protein Trapping**

The selected organosilica materials were placed in contact with a 0.5 mg/ml solution of lysozyme at pH 7.4 at room temperature during 4h, then they were filtered and the decrease on concentration of the protein in solution was measured by the UV adsorption at 280 nm. This allowed the calculation of the amount of lysozyme adsorbed per milligram of molecular sieve. As showed in Figure 5a, in the case of methyl-modified materials C18-M7-Me and C18-M15-Me the quantities adsorbed per milligram of organosilica (55 ± 1 µg/mg and 50 ± 5 µg/mg respectively) were lower than for the reference MCM-41 C18 (85 ± 10 µg/mg). Remarkably, in the case of the phenyl-modified sample, C18-M7-Ph, an adsorption 3.2 times higher was observed (270 ± 30 µg/mg).

Lysozyme adsorption is driven by electrostatic interactions between the Si-OH groups and the positively charged protein at pH 7.4. Therefore, the presence of –$CH_3$ instead of –OH groups on the surface could be one of the reasons of the lower quantity adsorbed by the methyl-modified silica. On the other hand, the high amount of lysozyme adsorbed by C18-M7-Ph is attributed to its ability to interact with the protein via electrostatic as well as π-π interactions.

FTIR spectra (Figure 5b) show peaks centered at 1656 cm$^{-1}$ and 1535 cm$^{-1}$ corresponding to the amide I and amide II bands of lysozyme, respectively. The small peak at 1594 cm$^{-1}$ comes from the aromatic C-C stretch of the phenyl functionalization. The amide I band consists of C=O stretching coupled with N-H bending and C-H stretching components. As these stretching modes are affected by the hydrogen bonds present in the different conformations (α-helix, parallel and antiparallel β-sheets, etc.), the shape of the amide I band gives information about the secondary structure of the protein[28]. It is then possible to perform studies of protein structure under different confined conditions using these new molecular sieves.

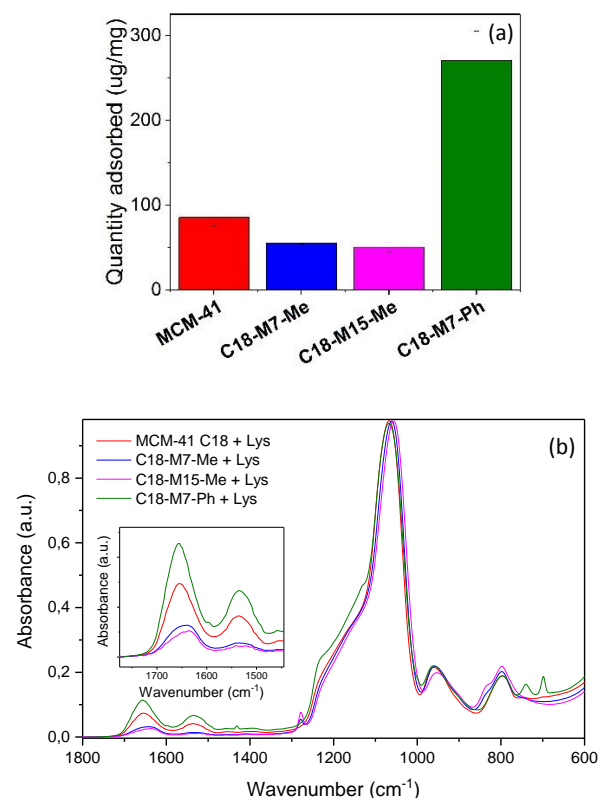

**Figure 5.** (a) Quantity of Lysozyme adsorbed from a 0.5 mg/ml solution at pH 7.4, per milligram of material (b) FTIR (ATR) Spectra of the lysozyme adsorbed on the materials. Insert: Zoom of the amide I and II bands of lysozyme.

**Catalysis of the alkyne-azide cycloaddition reaction**

The azide-alkyne cycloaddition reaction at ambient temperature, as shown in figure 6, typically requires a copper (I) species as catalyst. This species is prompt to disproportionate, making it necessary to add an auxiliary reducing species, as ascorbate, to have an efficient catalysis. Additionally, an extraction or filtration step has to be performed to purify the product[29]. Another alternative is to use copper loaded on nanoporous silica for heterogeneous catalysis.[12]

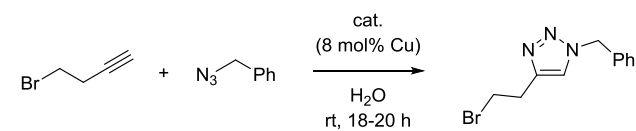

**Figure 6.** Cycloaddition reaction between bromobutyne and benzylazide in the presence of a copper catalyst in water.

Copper (II) was immobilized on the materials using the method developed by Brodie-Linder et al[30]. N$_2$ adsorption isotherms for the copper post-grafted materials (Figure 7) show a lower total amount adsorbed than the pristine samples.

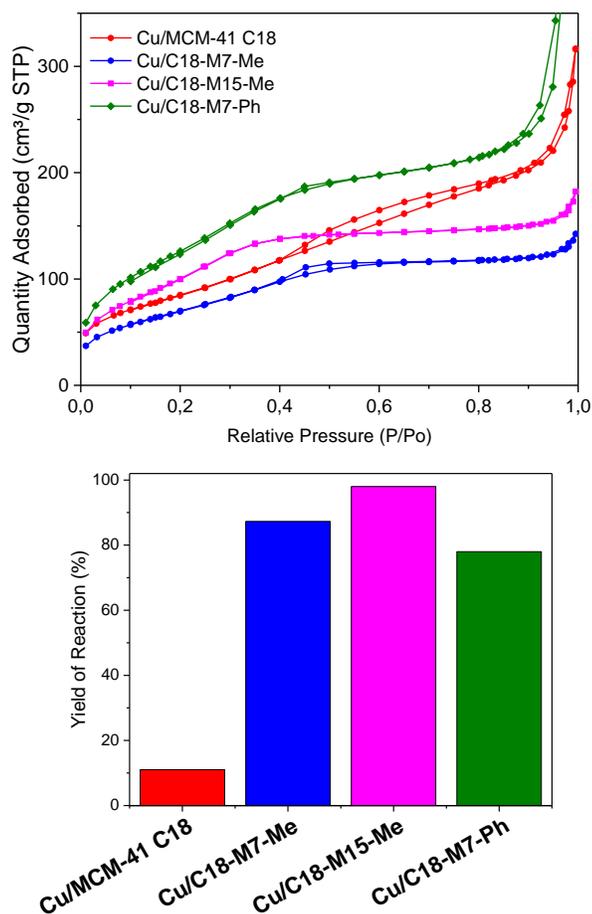

**Figure 7.** Nitrogen adsorption isotherms at 77 K (up) and yields of the cycloaddition reaction between bromobutyne and benzylazide in water (down) for the different silica-supported copper catalysts.

The irregular shape of the isotherm of Cu/MCM-41 C18 indicates more important wall degradation due to the contact with the basic copper solution. In this sense, the organic groups seem to protect the surface from basic hydrolysis, same as observed after water adsorption, Cu/C18-M7-Ph being the least affected in terms of surface area and pore volume. There is also an expected decrease on the pore size in all samples due to the incorporation of copper on the surface.

The materials were tested as catalysts for this reaction, with no addition of reducing agents or further ligands. Figure 7 shows the yields of reaction obtained by means of NMR spectroscopy. A vast increase in the yield of reaction is observed for the methyl- and phenyl- modified samples Cu/C18-M7-Me, Cu/C18-M15-Me and Cu/C18-M7-Ph (87%, >99% and 78% respectively) with respect to the yield of 11% obtained when Cu/MCM-41 C18 is used as catalyst.

These results confirm that the surface area and the organic functionalization of the materials play an important role in the efficiency of the catalyst. In this case, the presence of methyl or phenyl groups enhances the adsorption of the organic reactants. Another important factor is the pore diameter of the catalyst. It has been observed before that a tight fit that allows the entrance of the targeted molecules while also providing a confined environment for the reactants to interact, is a key factor in obtaining high reaction yields. Total transformation of the reactants on Cu/C18-M15-Me (pore size of 3.02 nm and a surface area of 377 m²/g) indicates that this material provides an optimum hydrophobic nanoenvironment for this particular reaction in water. A control experiment using copper-doped materials with pores blocked by surfactant (Figure S4) allowed confirmation that most of the conversion takes place inside the pores of the hydrophobically modified materials.

**Conclusion**

By means of a direct synthesis of TEOS with MTES or PTES, mesoporous organosilica materials were synthesized with different degrees of hydrophobicity for testing their applications in lysozyme trapping and supported copper catalysis. A complete characterization confirmed the large surface area, ordered pore structure and organic functionalization of the materials, providing a partial hydrophobic environment that depends on the percentage of organosilane present in the precursor mixture. It was observed that such a difference in the nanoenvironments had a significant impact on the performance of the materials in the selected applications.

The quantity of lysozyme adsorbed decreased for all the methyl-modified materials, with respect to the original MCM-41 C18. In contrast, the lysozyme uptake by the phenyl-modified sample, C18-M7-Ph, was more than three times higher than the reference MCM-41. This enhancement is attributed to π-π interactions between the organosilica and the protein.

The copper doped materials were proven to be active for regiospecific catalysis of the cycloaddition reaction between bromobutyne and benzyl azide, using water as solvent. In particular, C18-M15-Me has provided an ideal surface for this specific reaction, with an obtained yield of reaction of >99%. These unprecedented results for a Cu (II) supported catalyst, using such a facile synthesis encourages the testing of the materials for the synthesis of further triazole derivatives in water.

## ASSOCIATED CONTENT

**Supporting Information**

The Supporting Information is available free of charge on the ACS Publications website.
FTIR spectra, N$_2$ adsorption isotherms before and after water adsorption, TEM images and yields of catalysis reaction with blocked pores (PDF).

## AUTHOR INFORMATION

**Corresponding Authors**
*E-mail: christiane.alba-simionesco@cea.fr
*E-mail: nancy.brodie-linder@u-cergy.fr, nancy.linder@cea.fr
*E-mail: oriana.osta@cea.fr

## ACKNOWLEDGMENT

The authors thank Veronique Arluison for helpful discussion on the confined lysozyme, Annie Brulet for the SAXS exper-

## REFERENCES

(1) Lei, C.; Shin, Y.; Magnuson, J. K.; Fryxell, G.; Lasure, L. L.; Elliott, D. C.; Liu, J.; Ackerman, E. J. Characterization of Functionalized Nanoporous Supports for Protein Confinement. *Nanotechnology* 2006, *17*, 5531–5538.

(2) Charnay, C.; Bégu, S.; Tourné-Péteilh, C.; Nicole, L.; Lerner, D. A.; Devoisselle, J. M. Inclusion of Ibuprofen in Mesoporous Templated Silica: Drug Loading and Release Property. *Eur. J. Pharm. Biopharm.* 2004, *57* (3), 533–540.

(3) Cashin, V. B.; Eldridge, D. S.; Yu, A.; Zhao, D. Surface Functionalization and Manipulation of Mesoporous Silica Adsorbents for Improved Removal of Pollutants: A Review. *Environ. Sci. Water Res. Technol.* 2018, *4* (2), 110–128.

(4) Dosseh, G.; Xia, Y.; Alba-Simionesco, C. Cyclohexane and Benzene Confined in MCM-41 and SBA-15: Confinement Effects on Freezing and Melting. *J. Phys. Chem. B* 2003, *107* (26), 6445–6453.

(5) Azaïs, T.; Laurent, G.; Panesar, K.; Nossov, A.; Guenneau, F.; Sanfeliu Cano, C.; Tourné-Péteilh, C.; Devoisselle, J.-M.; Babonneau, F. Implication of Water Molecules at the Silica–Ibuprofen Interface in Silica-Based Drug Delivery Systems Obtained through Incipient Wetness Impregnation. *J. Phys. Chem. C* 2017, *121* (48), 26833–26839.

(6) Brodrecht, M.; Kumari, B.; Breitzke, H.; Gutmann, T.; Buntkowsky, G. Chemically Modified Silica Materials as Model Systems for the Characterization of Water-Surface Interactions. *Z. Für Phys. Chem.* 2018, *232* (7–8), 1127–1146.

(7) Wu, Q.; Hu, X.; Yue, P. L.; Zhao, X. S.; Lu, G. Q. Copper/MCM-41 as Catalyst for the Wet Oxidation of Phenol. *Appl. Catal. B Environ.* 2001, *32*, 151–156.

(8) Marcilla Gomis, A. Effect of Mesoporous Catalysts on the Mainstream Tobacco Smoke of 3R4F and 1R5F Reference Cigarettes. *Am. J. Chem. Eng.* 2015, *3* (1), 1–18.

(9) Mbaraka, I.; Shanks, B. Design of Multifunctionalized Mesoporous Silicas for Esterification of Fatty Acid. *J. Catal.* 2005, *229*, 365–373.

(10) Roy, S.; Chatterjee, T.; Pramanik, M.; Roy, A. S.; Bhaumik, A.; Islam, S. M. Cu(II)-Anchored Functionalized Mesoporous SBA-15: An Efficient and Recyclable Catalyst for the One-Pot Click Reaction in Water. *J. Mol. Catal. Chem.* 2014, *386*, 78–85.

(11) Rostamnia, S.; Doustkhah, E. Nanoporous Silica-Supported Organocatalyst: A Heterogeneous and Green Hybrid Catalyst for Organic Transformations. *RSC Adv.* 2014, *4*, 28238–28248.

(12) Jlalia, I.; Gallier, F.; Brodie-Linder, N.; Uziel, J.; Augé, J.; Lubin-Germain, N. Copper(II) SBA-15: A Reusable Catalyst for Azide-Alkyne Cycloaddition. *J. Mol. Catal. Chem.* 2014, *393*, 56–61.

(13) Wang, X.; Lin, K. S. K.; Chan, J. C. C.; Cheng, S. Direct Synthesis and Catalytic Applications of Ordered Large Pore Aminopropyl-Functionalized SBA-15 Mesoporous Materials. *J. Phys. Chem. B* 2005, *109*, 1763–1769.

(14) Yokoi, T.; Yoshitake, H.; Tatsumi, T. Synthesis of Amino-Functionalized MCM-41 via Direct Co-Condensation and Post-Synthesis Grafting Methods Using Mono-, Di- and Tri-Amino-Organoalkoxysilanes. *J. Mater. Chem.* 2004, *14* (6), 951.

(15) Sayari, A.; Hamoudi, S. Periodic Mesoporous Silica-Based Organic-Inorganic Nanocomposite Materials. *Chem. Mater.* 2001, *13*, 3151–3168.

(16) Shang, F.; Sun, J.; Wu, S.; Yang, Y.; Kan, Q.; Guan, J. Direct Synthesis of Acid–Base Bifunctional Mesoporous MCM-41 Silica and Its Catalytic Reactivity in Deacetalization–Knoevenagel Reactions. *Microporous Mesoporous Mater.* 2010, *134*, 44–50.

(17) Macquarrie, D. J. Direct Preparation of Organically Modified MCM-Type Materials. Preparation and Characterisation of Aminopropyl–MCM and 2-Cyanoethyl–MCM. *Chem. Commun.* 1996, No. 16, 1961–1962.

(18) Fowler, C. E.; Burkett, S. L.; Mann, S. Synthesis and Characterization of Ordered Organo–Silica–Surfactant Mesophases with Functionalized MCM-41-Type Architecture. *Chem. Commun.* 1997, No. 18, 1769–1770.

(19) Putz, A.-M.; Wang, K.; Len, A.; Plocek, J.; Bezdicka, P.; Kopitsa, G. P.; Khamova, T. V.; Ianăşi, C.; Săcărescu, L.; Mitróová, Z. Mesoporous Silica Obtained with Methyltriethoxysilane as Co-Precursor in Alkaline Medium. *Appl. Surf. Sci.* 2017, *424*, 275–281.

(20) Burkett, S. L.; Sims, S. D.; Mann, S. Synthesis of Hybrid Inorganic–Organic Mesoporous Silica by Co-Condensation of Siloxane and Organosiloxane Precursors. *Chem. Commun.* 1996, *11*, 1367–1368.

(21) Marei, N. N.; Nassar, N. N.; Vitale, G. The Effect of the Nanosize on Surface Properties of NiO Nanoparticles for the Adsorption of Quinolin-65. *Phys. Chem. Chem. Phys.* 2016, *18*, 6839–6849.

(22) Schoeffel, M.; Brodie–Linder, N.; Audonnet, F.; Alba–Simionesco, C. Wall Thickness Determination of Hydrophobically Functionalized MCM-41 Materials. *J Mater Chem* 2012, *22*, 557–567.

(23) Kruk, M.; Jaroniec, M.; Sayari, A. Adsorption Study of Surface and Structural Properties of MCM-41 Materials of Different Pore Sizes. *J. Phys. Chem. B* 1997, *101*, 583–589.

(24) Xin, C.; Zhao, N.; Zhan, H.; Xiao, F.; Wei, W.; Sun, Y. Phase Transition of Silica in the TMB–P123–H2O–TEOS Quadru-Component System: A Feasible Route to Different Mesostructured Materials. *J. Colloid Interface Sci.* 2014, *433*, 176–182.

(25) Frick, B.; Combet, J.; van Eijck, L. New Possibilities with Inelastic Fixed Window Scans and Linear Motor Doppler Drives on High Resolution Neutron Backscattering Spectrometers. *Nucl. Instrum. Methods Phys. Res. Sect. Accel. Spectrometers Detect. Assoc. Equip.* 2012, *669*, 7–13.

(26) Colmenero, J.; Mukhopadhyay, R.; Alegría, A.; Frick, B. Quantum Rotational Tunneling of Methyl Groups in Polymers. *Phys. Rev. Lett.* 1998, *80* (11), 2350–2353.

(27) Zorn, R.; Frick, B.; Fetters, L. J. Quasielastic Neutron Scattering Study of the Methyl Group Dynamics in Polyisoprene. *J. Chem. Phys.* 2002, *116* (2), 845–853.

(28) Adams, S.; Higgins, A. M.; Jones, R. A. L. Surface-Mediated Folding and Misfolding of Proteins at Lipid/Water Interfaces. *Langmuir* 2002, *18*, 4854–4861.

(29) Himo, F.; Lovell, T.; Hilgraf, R.; Rostovtsev, V. V.; Noodleman, L.; Sharpless, K. B.; Fokin, V. V. Copper(I)-Catalyzed Synthesis of Azoles. DFT Study Predicts Unprecedented Reactivity and Intermediates. *J. Am. Chem. Soc.* 2005, *127*, 210–216.

(30) Brodie-Linder, N.; Besse, R.; Audonnet, F.; LeCaer, S.; Deschamps, J.; Impéror-Clerc, M.; Alba-Simionesco, C. The Key to Control Cu II Loading in Silica Based Mesoporous Materials. *Microporous Mesoporous Mater.* 2010, *132*, 518–525.